\documentclass[reprint,twocolumn,secnumarabic,amssymb, nobibnotes, aps]{revtex4-1}

\usepackage{graphicx}
\usepackage{dcolumn}
\usepackage{bm}
\usepackage{amssymb}
\usepackage{latexsym}
\usepackage{amsmath}
\usepackage{appendix}

\begin{document}
\preprint{APS/123-QED}
\title{Role of flow of information in the speedup of quantum evolution}
\thanks{sps$\_$wuyn@ujn.edu.cn}
\author{Jing Wang,Yunan Wu$^*$,Ziyang Xie}
\affiliation{School of physics and technology, University of Jinan, Jinan, 250022, China}


\date{\today}

\begin{abstract}
The quantum evolution can be accelerated in non-Markovian
environment.\ Previous results showed that the formation of
system-environment bound state governs the quantum speedup. Although a
stronger bound state in the system-environment spectrum may seem like it
should cause greater speed of evolution, this seemingly intuitive thinking
may not always be correct. We illustrate this by investigating a
qubit driven by a classical field and coupled to a photonic crystal waveguide in
the presence of a mirror. The perfect mirror can force part of the emitted
light to return back to the qubit, and thus induce non-Markovian dynamics.
Within the considered model, we show how the evolution speed is influenced
by the memory time and the classical driving strength. In particular, we
find that the formation of bound state is not the essential reason for the
acceleration of evolution. The quantum speedup is attributed to the flow of
information, regardless of the direction in which the information flows. Our
conclusion can also be used to other non-Markovian environments.

\begin{description}
\item[PACS number(s)] 03.65.Ud,03.65.Yz,03.67.Mn
\end{description}
\end{abstract}

\maketitle

\preprint{APS/123-QED}

\affiliation{$^1$School of physics and technology, University of Jinan, Jinan, 250022, China}

\section{Introduction}

The quantum evolution speed determines how quickly a quantum system needs to
evolve between an initial state and a target state in a given process.
Realization of controllable speeding up of evolution of a quantum system
plays a key role in many technological applications, such as suppressing
decoherence  \cite{Georgescu153} and improving the efficiency of quantum computation \cite{Caneva240501,Lloyd1047}. For closed
quantum systems, it has been shown that the entanglement can accelerate the
quantum evolution \cite{Frowis052127,Batlr032337}. Due to the inevitable interaction between any system and
its environment, a considerable amount of work has witnessed research on
controlling speedup in more general open systems recently. One important
discovery is that the non-Markovian process induced by the memory
effect of environment can induce dynamical acceleration \cite{Defner010402,xu012307}, and therefore lead
to a smaller quantum speed limit time (QSLT), which is defined as the
minimal evolution time between two states \cite{Mandelstam249,Margolus188}.\ This phenomenon has been proved
by the experiment in cavity QED systems \cite{Cimmarusti233602}.

Much effort has been made to explore how to exploit the non-Markovian
environment itself to speed up quantum evolution.\ Some methods have been
provided to speed up quantum evolution for open systems, such as by
engineering multiple environments \cite{mo1600221}, driving the system by an external
classical field \cite{zhang032112}, and using the periodic dynamical decoupling pulse \cite%
{Song43654}.\ The reason of quantum speedup for the above methods is found to be
the increase of the degree of non-Markovianity. Recently, the authors of Re.~\cite{Liu020105}
showed that both the non-Markovianity and the quantum speedup are attributed
to the formation of system-environment bound state, i.e., the stationary
state of the whole system with eigenvalues residing in the band gap of
energy spectrum \cite{Yablonovitch2059,zhu2136,tong052330}. If the bound state is established, the evolution of system
becomes non-Markovian, and thus the quantum speedup happens. A good example
of this is the situation where a two-level atom is coupled to an environment
with a Ohmik spectrum. For this model, it has been found that providing
stronger bound states can lead to higher degree of non-Markovianity, and
hence to greater speed of quantum evolution \cite{Liu020105}. Based on this monotonic
relation between the three, controlling speedup through manipulation of
system-environment bound state has recently been studied \cite{behzzdi052121}. In some
sense, one may intuitively think that the formation of bound state can be
seen as the essential reflection to the speedup of quantum evolution.
However, the mechanism for quantum speedup in non-Markovian quantum systems
is still poorly understood if the environment is much complex.

The purpose of this paper is to examine the relationship between the
formation of bound state, non-Markovianity and the quantum speedup. To do
so, we consider a classical-driven atom coupled to a single-end
one-dimensional (1D) photonic crystal (PC) waveguide. The end of the PC
waveguide can be seen as a perfect mirror, forcing part of the emitted light
to return back to the atom. The feedback behavior may induce the information
backflow, i.e., the non-Markovian dynamics \cite{Breuer210401}. This
structure has been used to develop single-photon transistors \cite{Chang807}
and atomic light switches \cite{Zhou100501}. In this setting, the speedup
process for the embedded atom can be acquired by manipulation of the classical
driving field and the memory time of the environment. As for the mechanism
of quantum speedup, some unexpected and nontrivial results are found. The
formation of bound state can indeed lead to the non-Markovian evolution, but
does not necessarily result in quantum speedup. The speedup of quantum
evolution is attributed to the flow of information, regardless of the direction of information flows.
We illustrate that it is not the amount of
backflow information, i.e., the non-Markovianity, but the information flow volume
that ultimately determines the actual speed of quantum
evolution.

The work is organized as follows. The physical model is given in Section \uppercase\expandafter{\romannumeral2}.
In Section \uppercase\expandafter{\romannumeral3}, we construct the measure of actual speed of quantum evolution
based on information geometric formalism. Then we use this measure to
investigate how the environment affect the speed of quantum evolution within
our model in Section \uppercase\expandafter{\romannumeral4}. In order to clarify the mechanism for quantum
speedup, we first explore the interrelationship between the formation of
bound state, non-Markovianity and the quantum speedup in Section \uppercase\expandafter{\romannumeral5}, and then
present the role of the flow of information in the speeding up of evolution
in Section \uppercase\expandafter{\romannumeral6}. We summarize our results in Section \uppercase\expandafter{\romannumeral7}.

\section{Physical model}

We consider a qubit (two-level atom) with frequency $\omega _{0}$ driven by
a classical field with frequency $\omega _{L}$. The qubit is embedded in a
planar PC platform \cite{john2418,john12772} comprising a
semi-infinite 1D waveguide along $x$-axis (see Fig. 1). The
1D waveguides, whose end lies at $x=0$, are coupled to the driven qubit at $%
x=x_{0}$. By neglecting the counter-rotating terms, the Hamiltonian is ($%
\hbar =1$)
\begin{figure}[tbp]
\includegraphics[
height=2.0217in, width=2.8456in
]{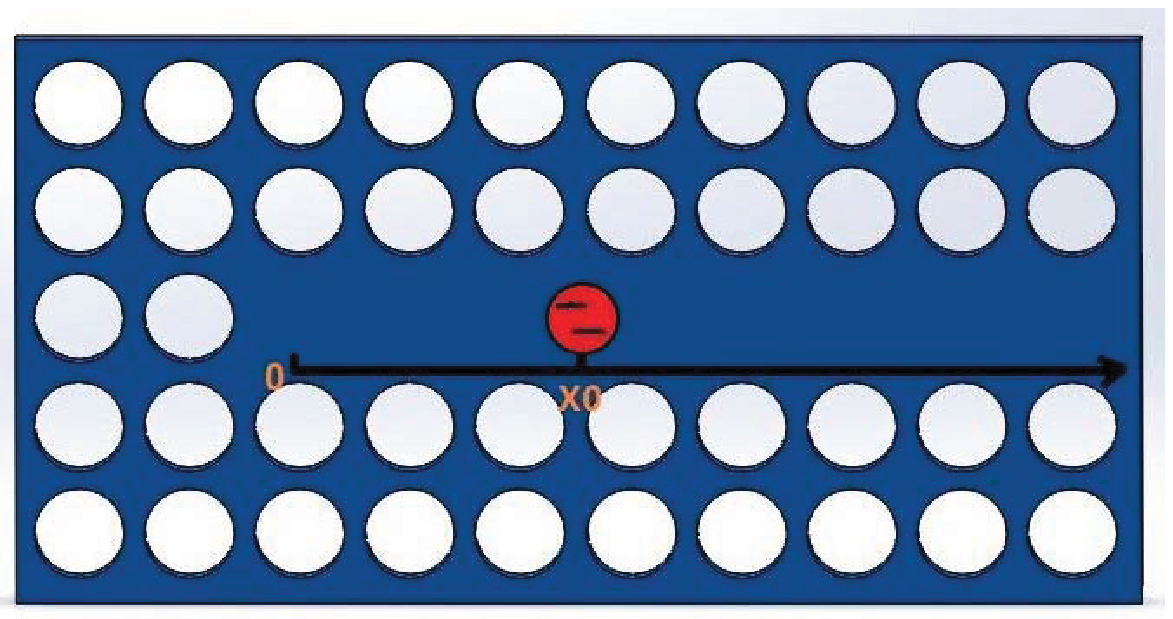}
\caption{The implementation of the model. A single-end PC waveguide, whose end lies at $x=0$, is coupled to a two-level atom (qubit) at $x=x_{0}$.}
\label{scabeh_sus}
\end{figure}
\begin{eqnarray}
&&H=\frac{\omega _{0}}{2}\sigma _{z}+\sum_{k}\omega _{k}a_{k}^{\dagger}a_{k} \notag \\
&&+\Omega \left( e^{-i\omega _{L}t}\sigma _{+}+H.c.\right)
+\sum_{k}(g_{k}a_{k}^{+}\sigma _{-}+H.c.),  \label{h1}
\end{eqnarray}%
where $\Omega $ is the coupling constant between the qubit and the classical
field, $\sigma _{z}=\left\vert e\right\rangle \left\langle e\right\vert
-\left\vert g\right\rangle \left\langle g\right\vert $ and $\sigma
_{+}=\sigma _{-}^{\dagger }=\left\vert e\right\rangle \left\langle
g\right\vert $ are the inversion and raising operators of the qubit with
excited and ground states $\left\vert e\right\rangle $ and $\left\vert
g\right\rangle $, $a_{k}$ ($a_{k}^{\dagger }$) is the annihilation
(creation) operator for the $k$th field mode with frequency $\omega _{k}$,
and $g_{k}$ is the coupling strength between the qubit and the $k$th mode.

The termination of the PC waveguide imposes a hard-wall boundary
condition on the field, and that the part of emitted photon of the qubit
will perform a round trip between the termination and the qubit. Thus the
waveguide termination can be considered a perfect mirror. This semi-infinite
1D structure can provide a broad quantum optical properties \cite%
{Tufarelli013820,Horak459}.

In this model, the photon dispersion relationship around the atomic
frequency is of the form \cite{Shen2001}
\begin{equation}
\omega _{k}=\omega _{0}+\upsilon (k-k_{0}),
\end{equation}%
where $\upsilon $ is the photon group velocity, and $k_{0}$ is the carrier
wave vector with $\omega _{k0}=\omega _{0}$. The coupling strength $g_{k}$
can be given by \cite{Tufarelli012113} \
\begin{equation}
g_{k}=\sqrt{\Gamma \upsilon /\pi }\sin kx_{0}
\end{equation}%
with the spontaneous emission rate $\Gamma $ of the qubit. For
convenience of calculations, we first give the effect Hamiltonian of our
model. By using the unitary transformation $U=\exp (-i\omega _{L}\sigma
_{z}t/2)$, the Hamiltonian in Eq. (\ref{h1}) can be transferred to
\begin{equation}
H=\frac{\Delta }{2}\sigma _{z}+\Omega \sigma _{x}+\sum_{k}\omega
_{k}a_{k}^{\dagger }a_{k}+\sum_{k}(g_{k}a_{k}^{+}\sigma _{-}e^{-i\omega
_{L}t}+H.c.),
\end{equation}%
where $\Delta =\omega _{0}-\omega _{L}$.\ In the basis $\{\left\vert +\right\rangle =\cos \frac{\eta }{2}\left\vert
e\right\rangle +\sin \frac{\eta }{2}\left\vert g\right\rangle ,\left\vert
-\right\rangle =-\sin \frac{\eta }{2}\left\vert e\right\rangle +\cos \frac{%
\eta }{2}\left\vert g\right\rangle \}$ with $\eta =\tan ^{-1}(2\left\vert
\Omega \right\vert /\Delta )$, the first
two terms on the right hand side of the above equation can be diagonalized,
and then the effect Hamiltonian can be rewritten
as
\begin{equation}
H_{eff}=\frac{\omega _{ef}}{2}\xi _{z}+\sum_{k}\omega _{k}a_{k}^{\dagger
}a_{k}+\sum_{k}(G_{k}a_{k}^{+}\xi _{-}+H.c.),  \label{he}
\end{equation}%
where $\omega _{ef}=\sqrt{\Delta ^{2}+4\left\vert \Omega \right\vert ^{2}}$,
$G_{k}=\cos ^{2}\frac{\eta }{2}g_{k}$, $\xi _{+}=\xi _{-}^{\dagger
}=\left\vert +\right\rangle \left\langle -\right\vert $ and $\xi
_{z}=\left\vert +\right\rangle \left\langle +\right\vert -\left\vert
-\right\rangle \left\langle -\right\vert $.

At zero temperature, let us consider that the qubit is in the state $%
\left\vert +\right\rangle $ and the reservoir in the vacuum state $%
\left\vert \tilde{0}\right\rangle $. By the Hamiltonian described in Eq. (%
\ref{he}), the state vector of the system at any time $t$, in the
interaction picture, can be given by
\begin{equation}
\left\vert \varphi \left( t\right) \right\rangle =c_{+}\left( t\right)
\left\vert +,\tilde{0}\right\rangle +\sum_{k}c_{k}\left( t\right) \left\vert
-,\tilde{1}_{k}\right\rangle , \label{cktt}
\end{equation}%
where the state $\left\vert \tilde{1}_{k}\right\rangle $ accounts for the
field mode with frequency $\omega _{k}$ having one excitation.\ By using the
Schr\H{o}inger equation, the amplitudes $c_{+}\left(
t\right) $ and $c_{k}\left( t\right) $ are governed by
\begin{eqnarray}
\dot{c}_{+}\left( t\right) =-i\sum_{k}G_{k}e^{i\left( \omega _{ef}-\omega
_{k}\right) t}c_{k}\left( t\right) ,  \label{c+} \\
\dot{c}_{k}\left( t\right) =-iG_{k}c_{+}\left( t\right) e^{-i\left( \omega
_{ef}-\omega _{k}\right) t}.  \label{ck}
\end{eqnarray}

By formal time integration of Eq. (\ref{ck}) and substituting this into
the Eq. (\ref{c+}), the amplitude $c_{+}\left( t\right) $ can be
transferred to
\begin{equation}
\dot{c}_{+}\left( t\right) =-\int_{0}^{t}dt^{^{\prime }}f(t-t^{^{\prime
}})c_{+}\left( t^{^{\prime }}\right)  \label{c11}
\end{equation}
with $f(t-t^{^{\prime }})=\underset{k}{\sum }\left\vert G_{k}\right\vert
^{2}e^{i(\omega _{ef}+\omega _{L}-\omega _{k})(t-t^{^{\prime }})}$. Through
integrating of the correlation function $f(t-t^{^{\prime }})$ over $k$ and
replacing this into Eq. (\ref{c11}), we acquire
\begin{eqnarray}
&&\dot{c}_{+}\left( t\right) =-\cos ^{4}\frac{\eta }{2}\frac{\Gamma }{2}%
c_{+}\left( t\right) \notag \\
&&+\cos ^{4}\frac{\eta }{2}\frac{\Gamma }{2}e^{i\left(
\omega _{ef}-\Delta \right) t_{d}}e^{i\phi }c_{+}\left( t-t_{d}\right)
\Theta \left( t-t_{d}\right),  \label{cdt}
\end{eqnarray}%
where $\Theta \left( t\right) $ is the Heaviside step function and $\phi
=2k_{0}x_{0}$.\  $t_{d}=2x_{0}/\upsilon $ is such that the finite time
taken by a photon to perform a round trip between qubit and the mirror,
which behaves as an environmental memory time \cite{Tufarelli012113}.
Obviously, the phase $\phi $ and the memory time $t_{d}$ are all dependent
on the atomic embedded position $x_{0}$, which can be accurately controlled
in the experiment \cite{Vats043808}. Performing the Laplace transformation
of Eq. (\ref{cdt}), we acquire
\begin{equation}
\tilde{c}_{+}\left( s\right) =\frac{1}{s+\cos ^{4}\frac{\eta }{2}\frac{%
\Gamma }{2}-\cos ^{4}\frac{\eta }{2}\frac{\Gamma }{2}e^{i(\omega
_{x}t_{d}+\phi )}e^{-st_{d}}},  \label{css}
\end{equation}%
where $\omega _{x}=\omega _{ef}-\Delta .$ By inverting the Laplace transform, we can obtain \cite{trung2524}
\begin{eqnarray}
c_{+}\left( t\right) =e^{-\cos ^{4}\frac{\eta }{2}\frac{\Gamma }{2}t}&\sum
\frac{1}{n!}(\cos ^{4}\frac{\eta }{2}\frac{\Gamma }{2}
e^{i\omega_{x}t_{d}}e^{i\phi }e^{\cos ^{4}\frac{\eta }{2}\frac{\Gamma }{2}%
t_{d}})^{n} \notag \\
&\cdot(t-nt_{d})^{n}\Theta \left( t-nt_{d}\right),
\end{eqnarray}
where the dynamical evolution is witnessed by the memory time $t_{d}$.

\section{Measure of dynamical speed}
The quantum speed of dynamical evolution can be constructed by applying the
method of differential geometry \cite{BengtssonBook}. Taking the perspective of this method,
the set of quantum states is indeed a Riemannian manifold, that is
the set of density operators over the Hilbert space. The geometric length
between the given initial state $\rho _{0}$ and the final state $\rho _{\tau
}$ can be naturally measured by using possible Reimannian metrics over the
manifold. According to the theorem of the Morozova, \v{C}encov and Petz
theorem \cite{Morozova2648,petz221,toth032324}, any monotone Riemannian metric can be given by the unified form
\begin{equation}
g^{f}\left( A,B\right) =\frac{1}{4}Tr\left( Ac\left( L_{\rho },R_{\rho
}\right) B\right) ,
\end{equation}%
where $A$ and $B$ are any hermitian operators, and $c\left( x,y\right) $
is a symmetric function defined as
\begin{equation}
c\left( x,y\right) =\frac{1}{yf(x/y)}
\end{equation}%
with $f(t)$ being the Morozova-\v{C}encov (MC) function which fulfills $%
f(t)=tf(1/t)$ and $f(1)=1$ \cite{Kubo205}.\ The MC function is related to our chosen
Riemannian metric, that is different forms of MC functions stand for
different Reimannian metrics. $L_{\rho }$ and $R_{\rho }$ are two linear
superoperators defined as $L_{\rho }A=\rho A$ and $R_{\rho }A=A\rho$.

Given the unified form of Riemannian metric, the squared infinitesimal
length between two neighboring quantum states $\rho $ and $\rho +d\rho $ can
be given by \cite{petz81}
\begin{equation}
ds^{2}=g^{f}\left( d\rho ,d\rho \right) .
\end{equation}
Here, we consider a dynamical evolution with a map $\Lambda _{t}$. The
evolved state is $\rho _{t}=\Lambda _{t}\rho _{0}$ with a initial state $%
\rho _{0}$. Along the evolved path between $\rho _{0}$ and $\rho _{\tau }$
with $t\in \lbrack 0,\tau ]$, the line element of the path can be expressed
as
\begin{equation}
dl=\sqrt{g^{f}\left( \partial _{t}\rho _{t},\partial _{t}\rho _{t}\right) }%
dt.
\end{equation}%
Then, the instantaneous speed of quantum evolution can be given by
\begin{equation}
V=\frac{dl}{dt}=\sqrt{g^{f}\left( \partial _{t}\rho _{t},\partial _{t}\rho
_{t}\right) }.
\end{equation}%
The average speed between time zero and $\tau $ is
\begin{equation}
V_{a}=\frac{1}{\tau }\int_{0}^{\tau }Vdt.
\end{equation}

In order to obtain the measure of dynamical speed in an explicit form, we
can rewrite the evolved state $\rho _{t}$ in the form of its spectral
decomposition, $\!\rho _{t}\!=\!\underset{k}{\sum }p_{k}\left\vert \phi
_{k}\right\rangle \left\langle \phi _{k}\right\vert\!$, with $0\!<\!p_{k}\!<\!1$ and $%
\underset{k}{\sum }p_{k}\!=\!1$. According to the Morozova-\v{C}encov-Petz
formalism, the instantaneous speed can be rewritten as \cite{BengtssonBook}
\begin{equation}
V=\sqrt{\underset{k}{\sum }\frac{\left\vert \dot{p_{k}}%
\right\vert ^{2}}{4p_{k}}+\underset{k\neq l}{\sum }c(p_{k},p_{l})\frac{%
p_{k}(p_{k}-p_{l})}{2}\left\vert \left\langle \phi_{l}|\dot{\phi_{k}}\right\rangle \right\vert ^{2}}.
\label{sulv}
\end{equation}
Clearly, any contractive Riemannian metric can be employed to evaluate the
speed of evolution with different type of MC function $f(t)$. As shown by
Re. \cite{Kubo205}, a generic MC function must fulfill $f_{\min }(t)<
f(t)< f_{\max }(t)$, where $f_{\min }(t)=2t/(1+t)$ and $f_{\max
}(t)=(1+t)/2$. Interestingly, an intermediate MC function with $f_{WY}(t)=(1+%
\sqrt{t})^{2}/4$ and $c_{WY}(x,y)=4/\left( \sqrt{x}+\sqrt{y}\right) ^{2}$ is
the one corresponding to the Wigner-Yanase information metric, which is
widely used in detecting the speed of dynamical evolution \cite{Deza2014}.\ In what follows,
we focus on the Wigner-Yanase information metric.\ Other potential appropriate
metric  is straightforward.

\section{\label{sec:level1}Controllable of quantum speedup}
In this section, we apply the measure constructed above to the 1D
waveguide system, and study the mechanism for controllable speedup. We
consider the atomic system is initially in an arbitrary pure state $%
\left\vert \Psi \left( 0\right) \right\rangle =\beta \left\vert
+\right\rangle +\sqrt{1-\beta ^{2}}\left\vert -\right\rangle $ $\left( 0\leq
\beta \leq 1\right) $, and the reservoir is in the vacuum state $\left\vert
\tilde{0}\right\rangle $. Exploiting Eq. (\ref{cktt}), the reduced density matrix for
the qubit can be calculated as (in the basis $\left\{ \left\vert
+\right\rangle ,\left\vert -\right\rangle \right\} $)
\begin{equation}
\rho _{a}\left( t\right) =\left(
\begin{array}{cc}
\beta ^{2}P_{t} & \beta \sqrt{1-\beta ^{2}}\sqrt{P_{t}} \\
\beta \sqrt{1-\beta ^{2}}\sqrt{P_{t}} & 1-\beta ^{2}P_{t}%
\end{array}%
\right) ,  \label{densitym}
\end{equation}%
where $P_{t}=\left\vert c_{+}\left( t\right) \right\vert ^{2}$ denotes the
excited state population of the qubit. The spectral decomposition of\ $\rho
_{a}\left( t\right) $ can be expressed as the form \
\begin{equation}
\rho _{a}\left( t\right) =\sum_{k=\pm }p_{k}\left\vert \phi
_{k}\right\rangle \left\langle \phi _{k}\right\vert ,  \label{atde}
\end{equation}
with $\!p_{\pm }\!=\!\left( 1\!\pm\! \lambda \right) /2\!$ and $\!\left\vert \phi _{\!\pm\!
}\right\rangle \!=\!\left( \alpha _{\pm }\left\vert\! +\!\right\rangle +\left\vert
-\right\rangle \right) /\sqrt{1\!+\!\alpha _{\pm }^{2}}$, where $\!\lambda \!=\!\sqrt{%
1\!-\!4\beta ^{2}P_{t}\!+\!4\beta^{4}P_{t}^{2}}$ and $\alpha _{\pm }\!=\!\left( 2\beta
^{2}P_{t}\!\pm\! \lambda \!-\!1\right) /\left( 2\beta \sqrt{1-\beta ^{2}}\sqrt{P_{t}\!}%
\right)$.\
\begin{figure}[tbp]
\includegraphics[
height=2.0217in, width=2.6456in
]{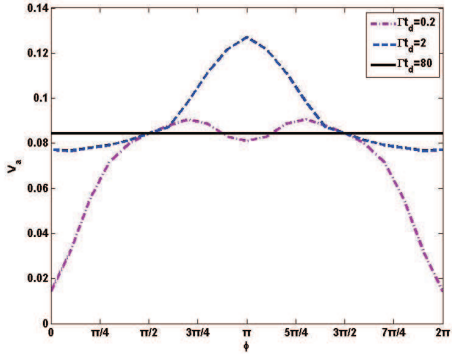}
\caption{The average speed of quantum evolution $V_{a}$ between the time zero and $\Gamma \tau =10$ (in units of $1/\Gamma $) as a function of phase $\phi$ for various values of memory time $\Gamma t_{d}$
with $\Omega=0$, $\Delta=0$ and $\beta=1$.}
\label{scabeh_sus}
\end{figure}
The dynamics of the qubit is fully determined by the Eq. (\ref{cdt}).
Clearly, the first term on the right side of Eq. (\ref{cdt}) is corresponding
to the atomic spontaneous emission.\ While the second term represents the
effect of the presence of the mirror on the atomic dynamical evolution.\  For
the sake of clarity, in what follows we consider three cases corresponding to the regimes of
small, intermediate and vary large values of $\Gamma t_{d}$, respectively.

\subsection*{\label{sec:level2}A. The small value of $\Gamma t_{d}$}

For simplicity, we first consider the case where there is no classical
field, i.e., the driving strength $\Omega =0$. Fig. 2 shows the average
speed between the time zero and $\Gamma \tau =10$ (in units of $1/\Gamma $)
for the system as a function of phase $\phi $. We can find that, in the regime where the memory time is small with
$\Gamma t_{d}=0.2$, the normalized
average speed for the qubit system is always relatively small. The maximum
value of $V_{a}$ is not exceed $0.1$ in the range $\phi \in \left[ 0,2\pi %
\right] $.

\begin{figure}[tbp]
\includegraphics[
height=2.0217in, width=2.6456in
]{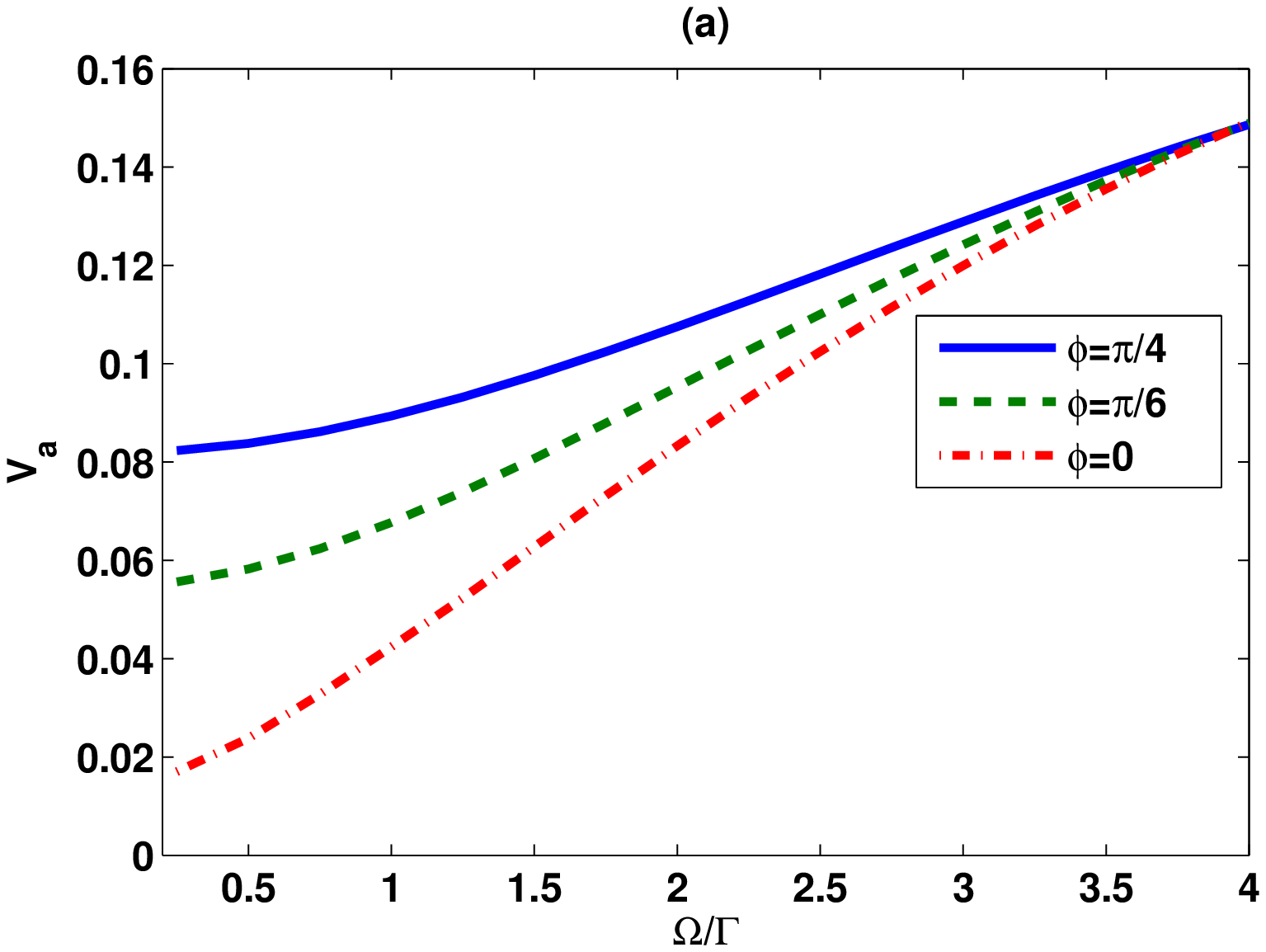} \includegraphics[
height=2.0217in, width=2.6456in
]{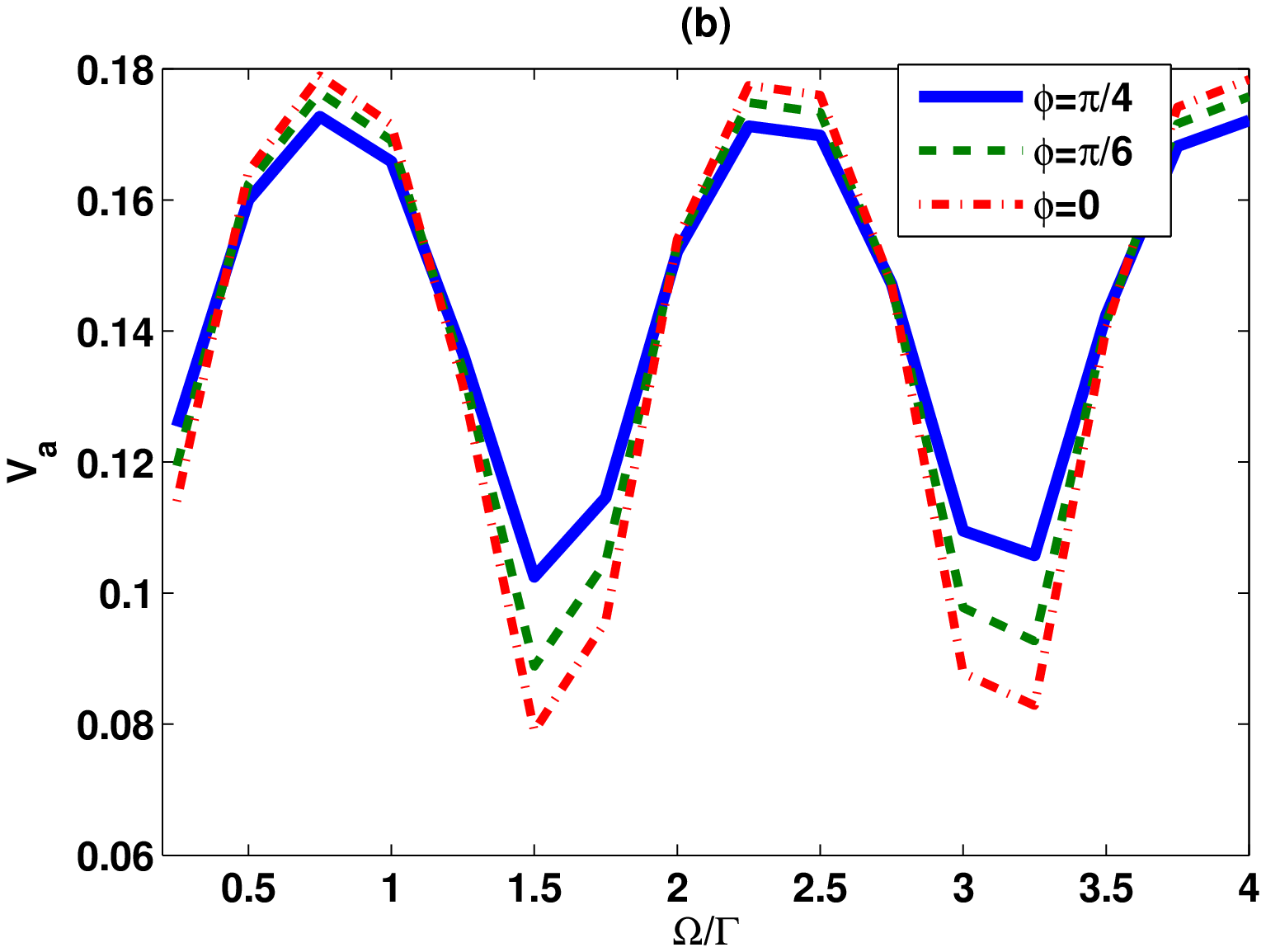}
\caption{The average speed of quantum evolution $V_{a}$ between the time zero and $\Gamma \tau =10$ as a function of classical driving strength $\Omega$ (in units of $1/\Gamma $)
 for various values of phase $\phi$ with $\Delta=0$ and $\beta=1$. (a) The memory time is small with $\Gamma t_{d}=0.2$. (b) The memory time is intermediate with $\Gamma t_{d}=2$.}
\label{scabeh_sus}
\end{figure}

In order to obtain the speedup of quantum evolution in this case, we show
how the classical field influence the speed of the qubit.  The variation of $V_{a}$ with respect to driving strength $%
\Omega $ for different $\phi $ is plotted in Fig. 3(a). For each line with
a fixed $\phi $, the increase of driving strength $\Omega $ leads to an
increase of the average speed. So, we therefore
reach the interesting result that the classical field can be used to speed
up the dynamical evolution in the case of the memory time is small.

\subsection*{B. The intermediate value of $\Gamma t_{d}$\ }

The regime of intermediate value of $\Gamma t_{d}$ means that the memory
time is shorter than the atomic spontaneous emission time, but can not be
ignored. In this situation, the qubit will undergo standard spontaneous
emission up to time $t=t_{d}$, which is independent of the phase $\phi$. After that, the presence of the mirror begins
to affect the dynamics of the open system.\ The fraction of light emitted by
the qubit will be reflected back to the qubit. So the presence of the mirror
is fully responsible for the non-Markovian character. In this case, the
speed of the atomic dynamical evolution is different from the above case, where the small memory time dose not induce any non-Markovianity \cite{Tufarelli012113}.
As shown from Fig. 2 (dashed line), over the entire range of the phase $\phi $,
the average speed in the intermediate regime ($\Gamma t_{d}=2)$ is bigger
than the case where the memory time is small. An interesting feature
here is that the speed $V_{a}$ has the obvious periodicity change under
action of the driving strength (as shown in Fig. 3(b)). That is to say, in the
intermediate regime, the speed of dynamical evolution can be controlled to a
speed-up and speed-down process by the classical driving strength.

\subsection*{C. The regime of very large value of $\Gamma t_{d}$\ }

It is worth noting the situation where the memory time is very large ($%
\Gamma t_{d}\gg 1$). In this case, the memory time is so large that the
emitted photon can not be reflected by the waveguide end, even when the
qubit has decayed to the ground state. Also, the dynamical evolution occurs
independently of the phase $\phi $ and the classical field. This is due to
the fact that the back-reflected light cannot recombine with the light
emitted towards the end of the waveguide and no interference takes place.
Thus, as expected, the average speed exhibits a plateau independent of $\phi
$, as shown in Fig. 2 (solid line).

In concluding this section, we would like to emphasize that, the classical
field as well as the the memory time $t_{d}$ play an important role in
controlling speedup. One can simply control the memory time (i.e., the
position of the embedded atom) and the driving strength to accelerate
dynamical evolution on demand. Our proposed scheme is experimentally
accessible. In experiment, the planar photonic crystal can be prepared by a
GaAs PC membrane, and the qubit can be prepared by the self-assembled InGaAs
QDs with a lower density \cite{Yoshie200}. We can use the method of electron
beam lithography to construct the photonic-crystal 1D waveguide.
Furthermore, the recent experiment has demonstrated an excellent control on
the atomic embedded position by the method of electrohydrodynamic jet printing \cite%
{See051101}. By this way, we can verify our prediction.

\section{Relationship between formation of bound state, non-Markovianity
and quantum speedup}

Previous result shows that \cite{Liu020105} the formation of the system-environment bound
state is the essential reason of the quantum speedup.\ It is demonstrated
that providing stronger bound states can lead to higher degree of
non-Markovianity, and hence to greater speed of quantum evolution.\ In order
to understand the physical reason of the speedup in our model, we further study the interrelation between the formation of bound
state, the non-Markovianity and the speed of quantum evolution.
\begin{figure}[tbp]
\includegraphics[
height=2.0217in, width=2.6456in
]{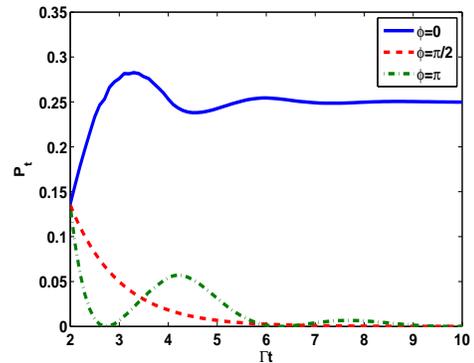}
\caption{The atomic excited population $P_{t}$ as a function of $\Gamma t$ for various values of
phase $\phi$ with $\Gamma t_{d}=2$, $\beta=1$, $\Delta=0$ and $\Omega=0$. Note that only the range $t>t_{d}$ is shown due to the fact that the behavior exhibits an exponential decay
and does not dependent on $\phi$ in $t<t_{d}$.}
\label{scabeh_sus}
\end{figure}

The system-environment bound state is actually the stationary state of the
whole system \cite{liu052139}. The formation of the bound state can lead to the inhibition
of spontaneous emission, i.e., the system holds an amount of excitation in
long time. Furthermore, the more stronger the bound state is, the
greater the amount of excitation bounded around the system is.\ This
phenomenon has been demonstrated in super-Ohmic \cite{Tong155501} and
photonic crystals bath \cite{John4083}. For our two-level atomic system,
the formation of bound state can be detected by the excited state population $P_{t}$ \cite{wu431},
which is sketched in Fig. 4. Obviously, it can be confirmed that, if $\phi
=\pi /2$, population decreases monotonically to zero, implying that the bound
state is absent.\ However, if $\phi =0,$ the population maintains a
steady-state value in long time limit.\ This population trapping behavior
means that fraction of light emitted by the qubit can only penetrate a
distance given by the length between the qubit and the mirror, and then be
reflected back to the qubit, forming the atom-photon bound state. While in
the case $\phi =\pi $, the bound state is established but it is not stronger
than the case that $\phi =0$. Thus, $P_{t}$ exhibits a periodically
decrease, that is only a small amount of excitation reflected back to the qubit
in this case.
\begin{figure}[tbp]
\includegraphics[
height=2.0217in, width=2.6456in
]{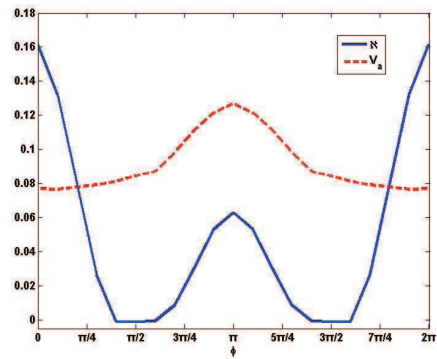}
\caption{The non-Markovianity $\aleph$ (solid line) and the average speed $V_{a}$ (dashed line) as a function of phase $\phi$
with $\Gamma t_{d}=2$, $\Omega=0$, $\Delta=0$ and $\beta=1$.}
\label{scabeh_sus}
\end{figure}

In Fig. 5, we sketch the behaviors of non-Markovianity $\aleph$ and the speed $V_{a}$.
The measure of non-Markovianity for our model is shown in the Appendix.
We can find that
the non-Markovianity connects directly with the formation of bound
state. When the bound state is established, such as in the case where $\phi=0$,  the system presents
non-Markovian effect with $\aleph>0.$ However, once the bound state is absent with $\phi
=\pi /2$, the behavior of sudden transition from non-Markovian to Markovian
effect ($\aleph=0$)  occurs. This result confirms the previous result that the
non-Marikovianity is attributed to the formation of bound states \cite{Liu020105}.

Now we focus on the speed of quantum evolution. When the bound state is
established and becomes stronger, the Markovian approximation of the
environment fails and one might expect the memory effect to accelerate the
speed of  evolution. This would be true if one were considering a simple
model where the qubit is directly connected to a reservoir taking Lorentzian
or Ohmic structures, as shown in Re \cite{behzzdi052121}. However, this relation may not be
universally true. When we considering our much complex physical model where
a classical driven qubit is confined in a controllable photonic waveguide,
a particularly astonishing phenomenon occurs. As shown in Fig. 5, when the
bound state is established with $\phi =0$ and the maximum non-Markovianity,
the average speed is $V_{a}=0.078$. While in the case where $\phi =\pi /2$ and $%
\aleph=0$ (Markovian effect), we can acquire $V_{a}=0.088$, which is bigger than the case, where the non-Markovianity is maximum in the range $\phi \in \left[ 0,2\pi %
\right] $.

Thus, the surprising message is that a stronger system-environment bound
state may not always helpful in enhancing the speed of quantum evolution.
What is the mechanism of speedup in a memory environment? What can be
seen as an essential reflection to the speedup of quantum evolution? To
answer the above questions, in the next section we further investigate the
speedup from the perspective of the direction of flow of information in memory
environments.

\section{Mechanism for the controlling speedup of quantum evolution}

The non-Markovian effect of environment connects tightly with the flow direction
of information. This is because the accepted notion of non-Markovianity
is based on the idea that the environment would cause the information
backflow from environment to the system for non-Markovian process, while for
Markovian process, the information flows in only one direction, that is from the
system to the environment, with no feedback \cite{Breuer210401}. The flow direction of information
can be monitored by the changing rate of the trace distance, i.e., $%
\sigma \left( t,\rho _{1,2}\left( 0\right) \right) =\frac{d}{dt}D\left( \rho
_{1}\left( t\right) ,\rho _{2}\left( t\right) \right) $.
The rate $\sigma \left( t,\rho _{1,2}\left( 0\right) \right) $ is positive
for an information backflow from environment to the system, and negative for
the information flowing in the opposite direction. Based on this, the total
amount of backflow information $\aleph=$ $\underset{\rho _{1,2}\left( 0\right) }{%
\max }\int_{\sigma >0}dt\sigma \left( t,\rho _{1,2}\left( 0\right) \right) $
is defined as the degree of non-Markovianity (see the Appendix).
\begin{figure}[tbp]
\includegraphics[
height=2.0217in, width=2.6456in
]{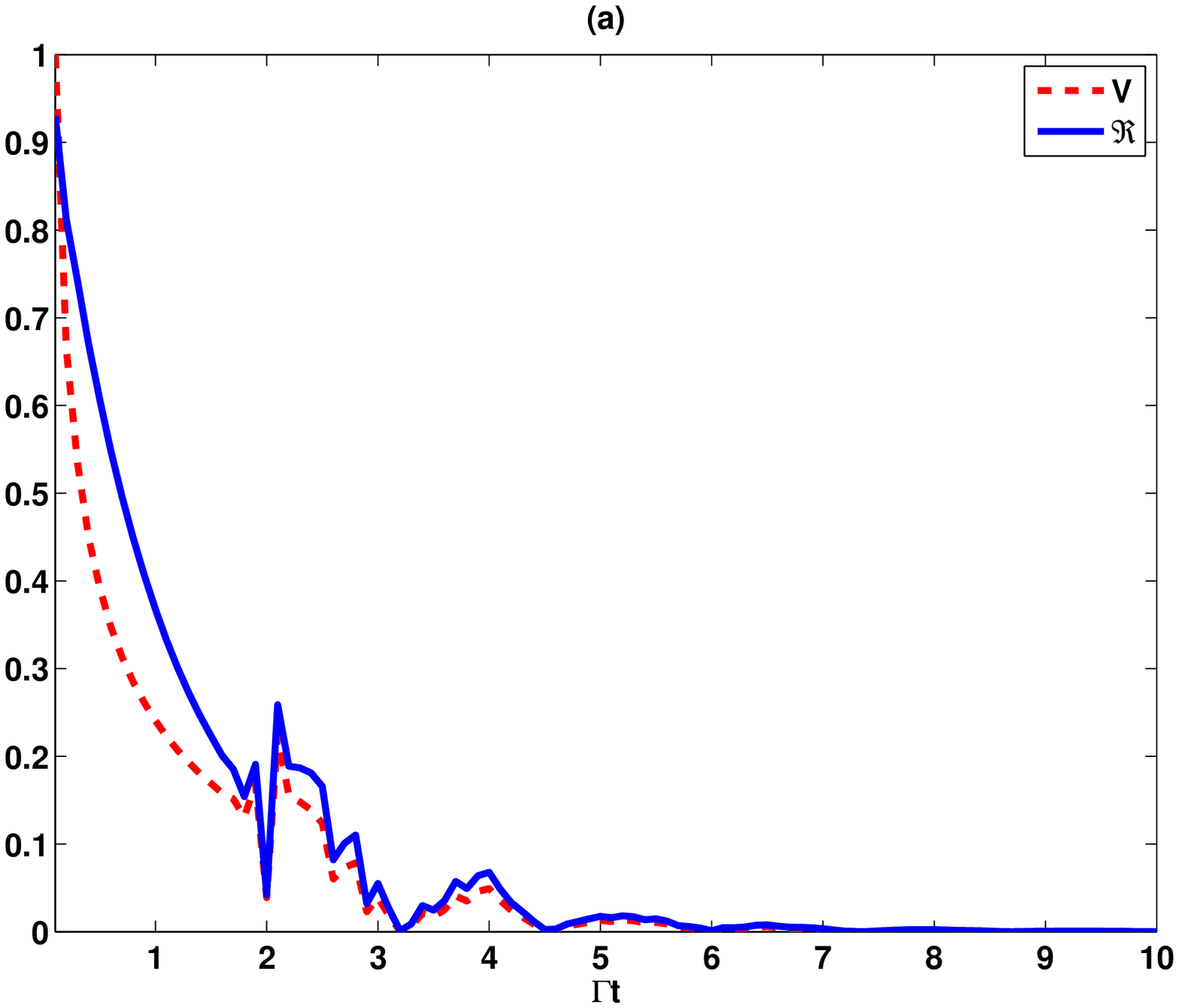} \includegraphics[
height=2.0217in, width=2.6456in
]{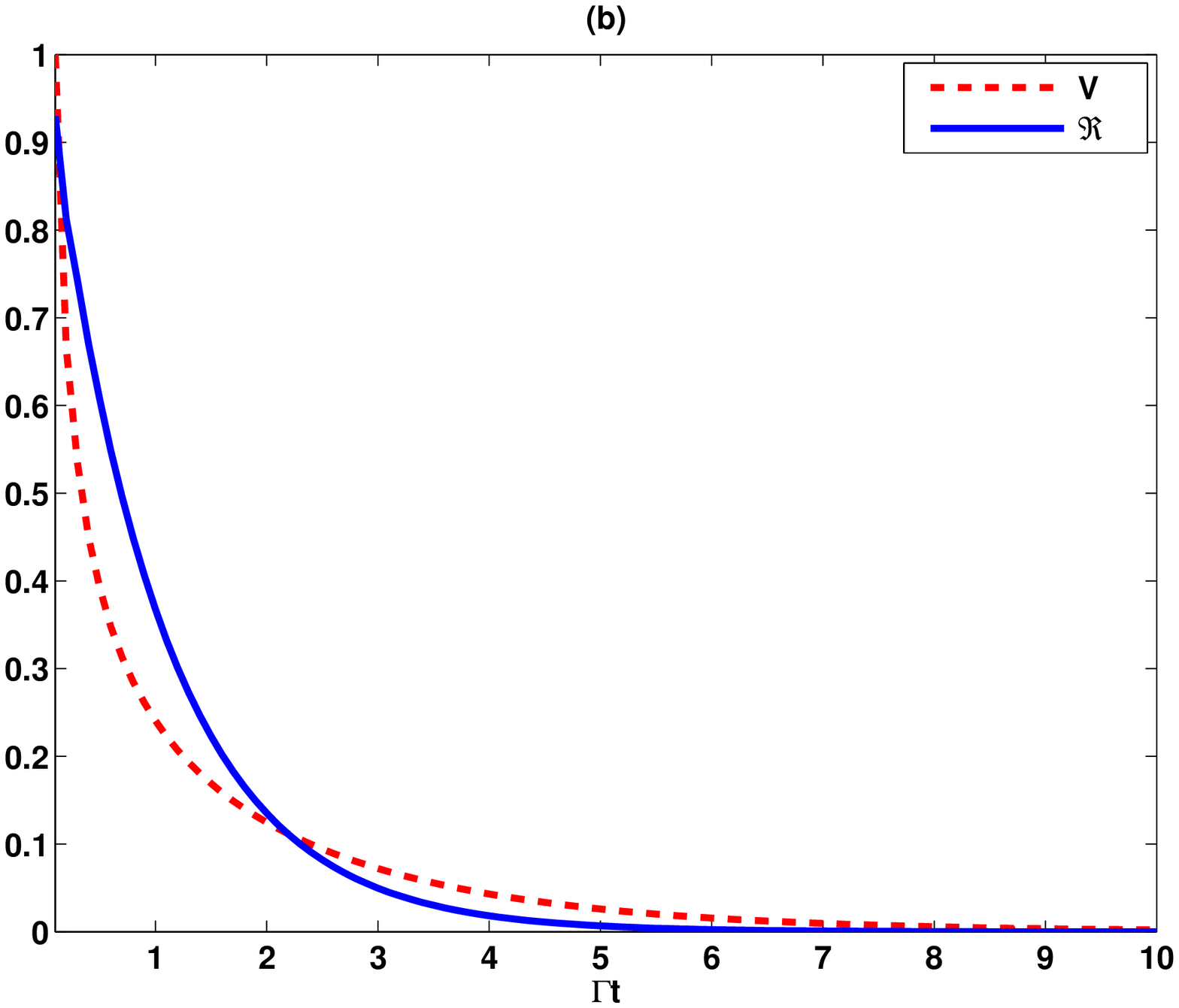}
\caption{The instantaneous speed $V$ (dashed line) and the absolute value of information changing rate $\Re= \left\vert \sigma \left(
t,\rho _{1,2}\left( 0\right) \right) \right\vert$ (solid line) as a function of $\Gamma t$ for (a) $\phi=0$ and (b)$\phi=\pi/2$ with $t_{d}=2, \Omega=0 and \Delta=0$.}
\label{scabeh_sus}
\end{figure}
Previous studies have shown that non-Markovian effect can speed up quantum
evolution. However, the degree of non-Markovianity $\aleph$ could not be seen as
an essential reflection to the quantum speedup. That is to say the reason
for the speedup is not solely to the backflow information. One question
naturally arise: What is the effect of the information flowing from system
to environment on quantum evolution?

Next, we focus on this question. In terms of above analysis, the total
amount of flow information  consisting the flow from system to environment
and the reverse flow is determined by
\begin{figure}[tbp]
\includegraphics[
height=2.0217in, width=2.6456in
]{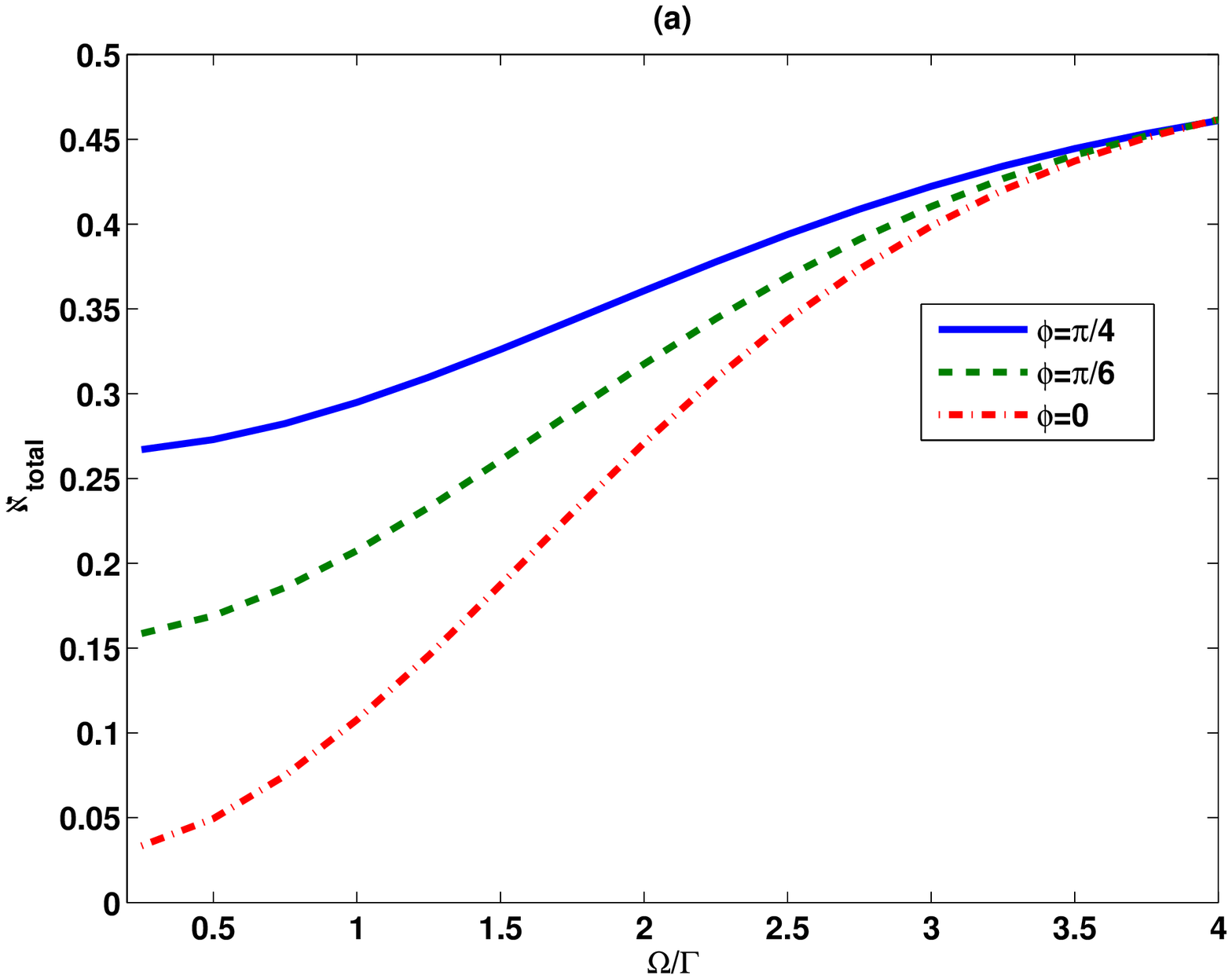} \includegraphics[
height=2.0217in, width=2.6456in
]{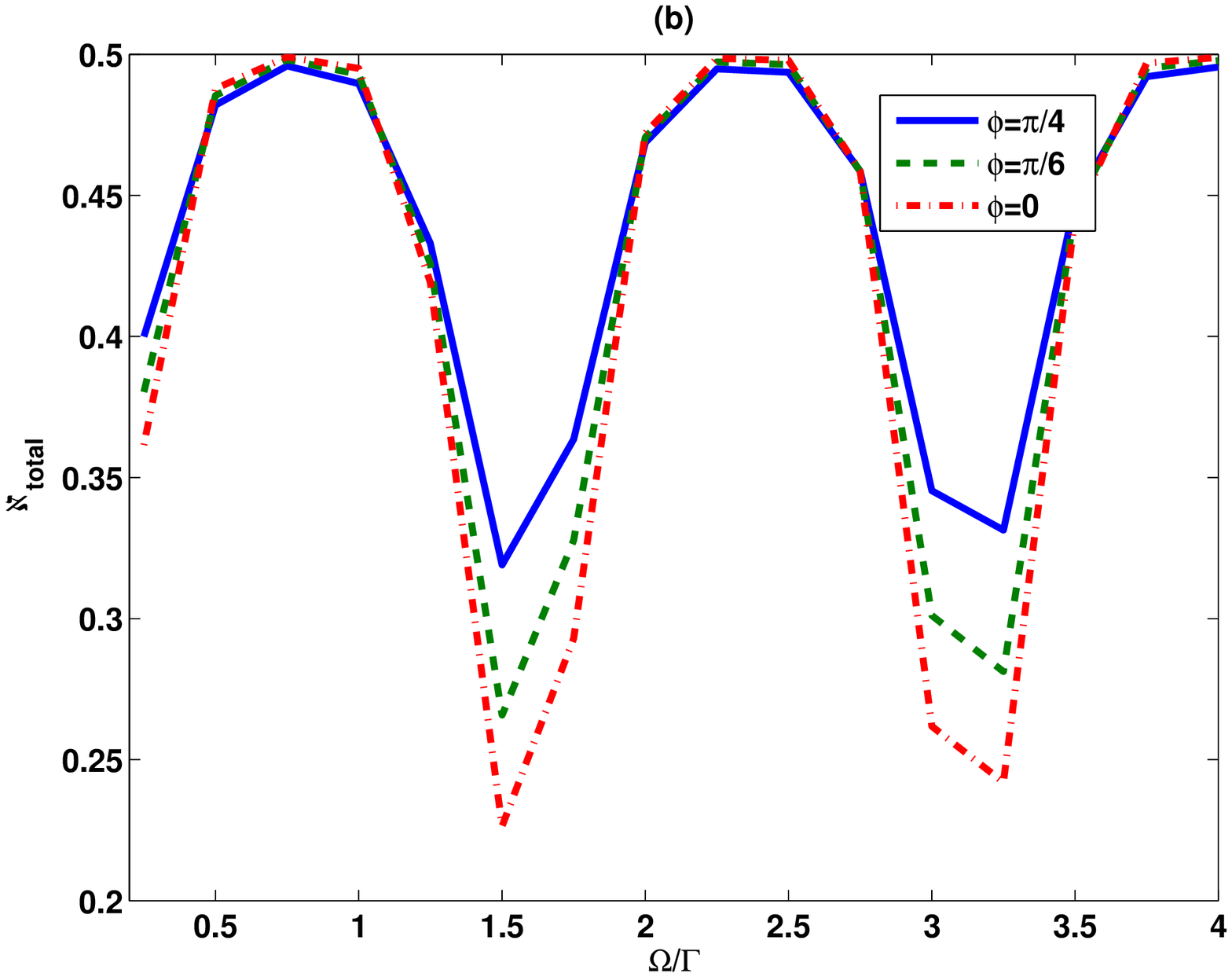}
\caption{The total amount of flow information $\aleph_{total}$ between the time zero and the time $\Gamma t=10$ as a function of classical driving strength $\Omega$ for various values of phase $\phi$ with $\Delta=0$. (a) The memory time is small with $\Gamma t_{d}=0.2$. (b) The memory time is intermediate with $\Gamma t_{d}=2$.}
\label{scabeh_sus}
\end{figure}
\begin{equation}
\aleph_{total}=\underset{\rho _{1,2}\left( 0\right) }{\max }\int dt\Re(t) ,
\end{equation}%
where the absolute value of changing rate $\Re(t)=\left\vert \sigma \left(
t,\rho _{1,2}\left( 0\right) \right) \right\vert $ denotes the flowing of
information. The comparison of the changing rate $\Re(t) $ and the instantaneous
speed $V$ for various values of phase $\phi $ with a fixed memory time is shown
in Fig. 6. It is interesting to find that the changing rate $\Re(t) $
exhibits the same behavior as the speed of quantum evolution. That is, the
increase (decrease) of $\Re(t) $ leads to an  increase (a decrease) of
instantaneous speed of quantum evolution. We thus conjecture that the flowing
of information plays a key role in controlling the speed of quantum
evolution. In order to further study the mechanism of controllable speeding
up of the evolution within the considered model, we plot in Fig. 7 the total
amount of flow information  $\aleph_{total}$ as a function of classical driving
strength $\Omega $ for various values of phase $\phi$. By contrasting the $%
\aleph_{total}$ and the average speed shown in Fig. 3, the results also confirm
that the driving strength can increase the information flow volume $\aleph_{total}$,
 and thus accelerate the quantum speed of evolution. We
therefore reach the interesting conclusion that it is the flow of
information that directly affects the quantum speed of evolution, regardless
of the direction of information flows. That is why in some cases, the Markovian
precess ($\aleph=0$) can also enhance the speed of evolution, as shown in Fig. 5.

\section{Conclusion}

In summary, we have studied a classical driven qubit that is coupled to an
1D photonic-crystal waveguide. We have investigated how the external
classical driving strength and the reservoir's memory time affect the
quantum speed of evolution.\ We find that, with a judicious choice of the
driving strength of the applied classical field, the speedup of evolution
can be achieved in both  Markovian (the memory time is small) and
non-Markovian (the memory time is intermediate) processes. We have also explored the
mechanism of well-controlled quantum speedup in our model. Surprisingly, a
stronger system-reservoir bound state with a higher degree of
non-Markovianity does not necessarily result in a greater speed of quantum
evolution.\ More specifically, within the considered model, we have shown
that it is not the amount of backflow information, i.e., the
non-Markovianity, but the total amount of flow information that directly
affect the average evolution speed for some interval of time.\ Our study sheds
further light on the interplay between information flowing and the evolution
speed of an open quantum system.

Finally, it should be note that our conclusion applies not only to the above
model, but also to situations like a qubit coupled to a environment with
Lorentzian or Ohmic structure.\ Within these models, it is easy to check
that population trapping will not be happened \cite{Defner010402,Liu020105,behzzdi052121}. The population $P_{t}$
exhibits a monotonic decay (Markovian dynamics) or a periodically decrease
(mon-Markovian dynamics). The difference of speed between the Markovian and
non-Markovian cases, i.e., the difference of the total amount of flow information ($\aleph_{total}$)
 between them, is mainly determined by the backflow information.\ The effect of the information flowing
 from system to environment can be ignored. Thus,
the non-Markovianity becomes the unique reason for quantum speedup in these
models.
\section{Acknowledgements}

This work was supported by the the Young Foundation of Shan Dong province
(Grant number ZR2017QA002) and Doctoral Foundation of University of Jinan
(Grant no.~XBS1325).

\appendix

\section{Measure  of non-Markovianity}

In non-Markovian dynamics, the environment would cause the information
backflow from environment to the system. The non-Markovianity describing the
total amount of backflow information is defined as
\begin{equation}
\aleph=\underset{\rho _{1,2}\left( 0\right) }{\max }\int_{\sigma >0}dt\sigma
\left( t,\rho _{1,2}\left( 0\right) \right) ,
\end{equation}
where $\sigma \left( t,\rho _{1,2}\left( 0\right) \right) =\frac{d}{dt}%
D\left( \rho _{1}\left( t\right) ,\rho _{2}\left( t\right) \right) $ denotes
the changing rate of the trace distance $D\left( \rho _{1}\left( t\right)
,\rho _{2}\left( t\right) \right) =\frac{1}{2}tr\left\vert \rho _{1}(t)-\rho
_{2}(t)\right\vert $ between states $\rho _{1,2}(t)$ evolving from their
respective initial states $\rho _{1,2}\left( 0\right) $ \cite{Breuer210401}%
. The dynamical process is non-Markovian if there exists a pair of initial
states and at certain time such that $\sigma \left( t,\rho _{1,2}\left(
0\right) \right) >0$. For our two-level system, the optimal pair of initial
states has been proven to be $\rho _{1,2}\left( 0\right) =\left\vert \pm
\right\rangle \left\langle \pm \right\vert $ \cite{Wissmann062108}. Then the
trace distance can be acquired $D\left( \rho _{1}\left( t\right) ,\rho
_{2}\left( t\right) \right) =\left\vert c_{+}\left( t\right) \right\vert
^{2} $.

\end{document}